\documentclass[authoryear,preprint,review,12pt]{elsarticle}

\usepackage{amssymb}
\usepackage{mathptmx}
\usepackage{nicefrac, xfrac}
\usepackage[margin=3cm]{geometry}
\usepackage{lineno}
\usepackage{tabularx}
\usepackage{xcolor}
\usepackage[hidelinks]{hyperref} 
\usepackage{ulem}
\hypersetup{
    colorlinks= true, %Colours links instead of ugly boxes
    linkcolor={red!50!black}, %Colour of internal links
    citecolor={blue!50!black},
    urlcolor={blue!80!black}
}
\journal{\ }
\begin{document}
\begin{frontmatter}
%% title to be improved !!!
\title{Efficient mechanical evaluation of railway earthworks using a towed seismic array and Bayesian inference of MASW data}

%% Authors currently ordered by teams working on the project + helps
% final order to be defined later
\author[1,2]{A.~Burzawa}\corref{cor1}
% team SncfReseau-SorbonneUniv
\author[1]{L.~Bodet}
\author[2]{M.~Dangeard}
% team ZETICA 
\author[3]{B.~Barrett}
\author[3]{D.~Byrne}
\author[3]{R.~Whitehead}
\author[3]{C.~Chaptal}

% help on the field 
\author[1,2]{J.~Cunha Teixeira}
\author[1]{J.~C\'{a}rdenas}
\author[1,4]{R.~Sanchez Gonzalez}

\author[3]{A.~Eriksen}
\author[2]{A.~Dhemaied}

%Conceptualization /Data curation / Formal analysis / Funding acquisition / Investigation / Methodology / Project administration / Resources / Software / Supervision / Validation / Visualization / Writing – original draft / Writing – review and editing

\address[1]{Sorbonne Université, CNRS, EPHE, UMR 7619 METIS, 4 place Jussieu, 75252 Paris 05, France}
\address[2]{SNCF Réseau, 6 avenue François Mitterrand 93210 Saint-Denis, France}
\address[3]{Zetica Ltd, Zetica House, Southfield Road, Eynsham, Oxford OX29 4JB, UK}
\address[4]{Geosciences Department, Mines Paris - PSL, PSL University, Paris, France}

\cortext[cor1]{audrey.burzawa@sorbonne-universite.fr}
\begin{abstract} %max 100 words (here 100)
Assessing Railway Earthworks (RE) requires non-destructive and time-efficient diagnostic tools. This study evaluates the relevance of shear-wave velocity ($V_s$) profiling using Multichannel Analysis of Surface Waves (MASW) for detecting Low Velocity Layers (LVLs) in disturbed RE zones. To enhance time-efficiency, a towed seismic setup (Landstreamer) was compared with a conventional one. Once qualified, the Landstreamer was deployed on the ballast for roll-along acquisition, showing greatly improved efficiency and good imaging capability. A probabilistic framework adopted in this study additionally enhances quantification of uncertainties and helps in interpretation of $V_s$ models, facilitating reliable decision-making in infrastructure management.
%Additionally, the combined interpretation of MASW and Ground Penetrating Radar (GPR) data confirms their strong complementary nature, providing a more comprehensive assessment of RE conditions.
\end{abstract}
% \begin{highlights}
% \item Need for non-destructive diagnosing tools for railway earthwork
% \item Contribution of shear-wave velocity profiling (MASW) along the cess
% \item Comparison of conventional and towed seismic acquisition setups
% \item Improving acquisition yield with a towed seismic setup on the track
% \item Benefit from the complementary nature of MASW and Ground Penetrating Radar (GPR) data
% \end{highlights}
\end{frontmatter}
%
%\linenumbers
%\newpage
%\tableofcontents
%\newpage

%%%
\section*{Acknowledgements}
\label{sec:acknowledgments}
%%%
This work has been funded by \textit{SNCF Réseau/CNRS/Université Pierre et Marie Curie-Paris6/Sorbonne Université} research contracts C13/411, C15/0781 and the convention \textit{ANRT/ Cifre-SNCF Réseau} research contracts C21/1995. This work has been funded by the European Union. Views and opinion expressed are however hose of the author(s) only and do not necessarily reflect those of the European Union or the granting authority. Neither the European Union nor the granting authority can be held responsible for them. The project is supported by the Europe’s Rail Joint Undertaking and its members. This project has received funding for them. The project is supported by the Europe’s Rail Joint Undertaking and its members. This project has received funding from the European Union’s Horizon Europe research and innovation programme under Grant Agreement No 101101966. The geophysical equipment was provided by the METIS laboratory at \textit{Sorbonne Université}. The DCP archive data was supplied by SolSolution and the GPR archive data by Ground Control. Seismic data processing has been performed thanks to open-source software packages: SWIP (\url{https://github.com/spasquet/SWIP/releases}), BayesBay for inversion \url{https://github.com/fmagrini/bayes-bay} and PAC (\url{https://github.com/JoseCunhaTeixeira/PAC}). The authors are grateful for the valuable collaboration with Zetica Ltd for their advices and support during the study, as well as valuable discussion for the results interpretations. The authors would like to thank the SNCF Réseau team and in particular M.~Fonda, A.~Hallier and J.~Boisson-Gaboriau for their help in the field and during data interpretation.

\section*{Introduction}
\label{sec:introduction}
The stability of railway tracks is of paramount importance to all those involved in rail network maintenance, particularly in relation to the mechanical quality and bearing capacity of railway trackbed (RT) and shallow railway earthwork (RE) \citep{Selig1994,putallaz2007HDR,Tzanakakis2013}. The current tools routinely used to characterise the mechanical quality of these soil layers are core sampling and Dynamic Cone Penetration (DCP) tests \citep{Escobar2016,Haddani2016}. These measurements provide strong localised knowledge of structure and quality, but are costly and time-inefficient when applied to the whole network. It is therefore essential to have non destructive diagnostic tools adapted to these issues. Geophysical methods meet the need for efficiency. However, the railway context makes the application of certain techniques particularly complex \citep{Artagan2020, Kyrkou2022}, and their use is often limited to geometric characterization \citep[][for more details about these issues]{Burzawa2023}

Fortunately, the mechanical characterisation of RT and shallow RE can also be achieved through the evaluation of the shear-wave velocity ($V_s$), which is directly related to the small-strain shear modulus ($G_0$ or $G_{max}$) \citep[][]{Dhemaied2014a, Byun2019, Burzawa2023}. Such soil parameters can be estimated \textit{in situ} through seismic methods \citep{Mari2004}. However, the most frequently used seismic refraction and reflection are ill-suited to the study of the RE due to their high sensitivity to three-dimensional (3D) effects and their difficulty in detecting low-velocity layers (LVLs) \citep{Ryden2004, 0neill_Matsuoka2005}. This type of structure is particularly common when the near-surface soil layers are highly compacted. However, surface-wave analysis has shown that they are less sensitive to 3D effects \citep{Karl2011} and enable LVLs to be identified \citep{Ryden2004}. Surface-wave methods have actually shown the efficiency for determining $V_s$ and stiffness profiles with depth of near-surface anthropogenic structures \citep{Foti_etal2014_Book}. The study of surface waves, and in particular Rayleigh waves, for small-scale applications and the determination of soil damping, has been extensively researched in civil engineering contexts \citep[][]{Lai1998,rix_eta2001simultaneous,Abbas2024}. A number of authors have for instance demonstrated the use of surface waves for the characterisation of pavements and concrete structures \citep{heisey1982moduli,Nazarian1984,Matthews1996,Hevin1998,Park_etal1999,Lai2002}. 

The most commonly employed method for the surface-wave analysis is the Multichannel Analysis of Surface Waves (MASW) \citep{Park_etal1999,xia1999estimation}. MASW is employed to retrieve one-dimensional (1D) $V_s$ profile with depth \citep{Socco2004,Foti2018} and is also used for profiling to provide two-dimensional (2D) $V_s$ image of shallow layers \citep{Socco2010,Pasquet2017}. Over the past decade, MASW has been employed in the railway context to characterise the RE \citep[][]{Donohue2013,Donohue2014,Hwang2014,Forissier2015,Gunn2015,Bergamo2016a,Bergamo2016b,Gunn2016,SussmannJr2017,Artagan2020,Kyrkou2022}. Previous studies have adapted seismic acquisition strategy to the scale of shallow RE and demonstrated high resolution on the application along a track maintenance anomaly zone \citep{bodet2017_SNCF_interne,Burzawa2023, Cunha_seismica2025}. 

Traditionally, seismic acquisition consists of planting sensors (called geophones) with spikes inserted into the ground \citep{Gunn2016,Burzawa2023}. One significant challenge is that the conventional method of planting a large number of geophones is particularly unsuitable for the purpose of profiling. Indeed, the process of profiling involves the deployment of seismic arrays along an extensive length, with the objective of encompassing a large zone (several hundred meters to several kilometres long). The deployment and data acquisition process with the conventional setup is then time-consuming, requiring several hours to complete for a distance of several tens of linear meters. To reduce acquisition time and adapt to the constraints of the railway context, this study proposes the implementation of a towed seismic setup, such as a Landstreamer (LS) \citep{inazaki1999_LS, inazaki2004high, Liberty2014}.

A seismic LS is a linear array of geophones connected by sturdy cables that are simply laid on the ground (so there is no need to plant them) and can be towed by a machine such as truck, trolley or lorry \citep{inazaki1999_LS}. The LS is employed in applications that necessitate a high acquisition yield while ensuring the accuracy of the data acquired \citep{Oneill_etal2006}. The use of LS for near-surface applications, particularly paving, has been investigated by numerous research teams \cite[e.g. pioneering applications by][]{VanDerVeen_Green1998,VanDerVeen1999,VanderVeen2001,inazaki2004high}. Additionally, LS has been used in civil engineering applications, such as the construction of dykes \citep{hayashi2006surface} and in urban areas \citep{inazaki2005high,suarez2007field,Liberty2014}. A large number of LS have been developed and have contributed their specific features and technological advances. Notably, recent developments have resulted in the advent of a LS equipped with DAS technology \citep{Pandey_etal2023_Developping}, with multicomponent broadband
MEMS (micro-electro mechanical system) geophones \citep{Brodic_etal2015, Brodic2018}. Furthermore, \cite{Liberty2024} introduced portable ‘hand streamers’ designed for rapid seismic imaging of shallow subsurface targets. The studies conducted by \cite{Oneill_etal2006, suarez2007field} and more recently \cite{Hanafy2022}, have investigated reworked soils beneath road infrastructures at spatial scales typically to the upper 5~m, which are comparable to those of RT and shallow RE surveys. These investigations employed a fine geophone spacing of 0.5~m to ensure adequate resolution. \cite{Oneill_etal2006} and more recently \cite{Cyz2024} showed by comparison of data obtained from LS and conventional seismic setups that the two data sets were comparable. In the railway context, LS has been deployed on the cess in order to characterise RE \cite[][]{Gunn2016} but not commonly on ballast and along long linear sections. This study evaluates the effectiveness of a LS for deployment in a railway context, with the aim of providing $V_s$ imaging along a specific section of track.

In addition, this study adopts the probabilistic approach to provide a practical framework for surface-wave dispersion inversion, . Instead of relying on fixed parameter estimates with associated uncertainties, variations of parameters of the inverse problem are represented as probability distributions. This approach, allows a fairer assessment of $V_s$ and thickness $H$ spatial changes along the railway track. The Bayesian framework \citep{Tarantola2005} enables this by quantifying uncertainty and integrating \textit{prior} information, particularly high in areas where geotechnical and geophysical investigations are conducted. This approach has been applied to large-scale surface-wave dispersion inversion \citep{Sambridge1999a, Sambridge1999b, Bodin2012} and, more recently, to near-surface investigations, where it enables the integration of various geophysical and geotechnical data types \citep{Killingbeck2018, Hallo2021}. In the railway context, \cite{Burzawa2023} have previously demonstrated the value of Bayesian methods for estimating shear-moduli contrasts. 

In this study, a specific site along the South East European High-Speed-Lines (HSL) was subjected to a comprehensive investigation. The site exhibited track geometry disorders due to trackbed failure. This investigation was conducted through the implementation of geotechnical and geophysical surveys. MASW with a conventional setup (geophones planted on the cess) was first used to distinguish a structural and/or mechanical anomaly in the RE with $V_s$ vertical profiles, along the anomaly zone. MASW acquisitions were actually carried out on the cess at two distinct positions with respect to the localisation of the anomaly (called P0 and P1 in the rest of this study). These same two locations were selected and investigated again, this time to validate the applicability of a LS, with a strict comparison between the data acquired with the conventional setup. Once the use of the LS setup validated on the cess, profiling was carried out on the track using the LS over a 300~m long zone. The inversion process is finally presented using Bayesian formalism to extract reliable statistical estimators relevant to infrastructure managers.  

\section*{Site description}
\label{sec:site_description}
\subsection*{Context of track maintenance anomalies}
\begin{figure}
    \includegraphics[width=0.9\linewidth]{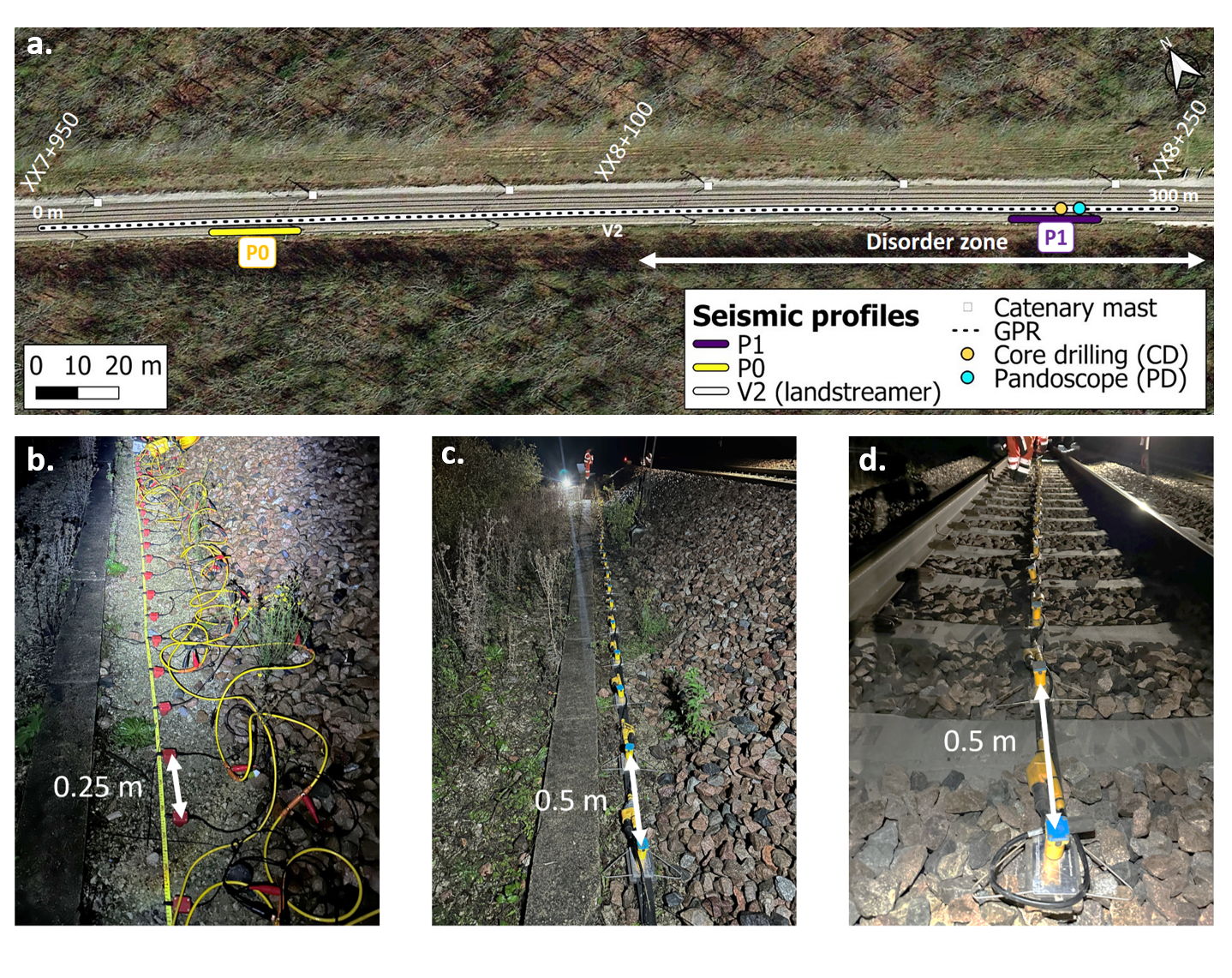}
    \centering
    \caption{\textbf{a.} Schematic map of the study site. Positions of geotechnical surveys (core drilling and DCP) and geophysical acquisitions (seismic and GPR) are shown. The position correspond to Kilometre Point (KP XXX+YYY) whose full position is confidential. \textbf{b.} Photography of the conventional seismic setup placed on the cess. \textbf{c.} Photography of the landstreamer placed on the cess. \textbf{d.} Photography of the landstreamer placed on the track.
    }
    \label{fig:map}
\end{figure}

The study area is located on the South-East European HSL between Paris and Lyon (Figure~\ref{fig:map}a). This was the first HSL to be built and put into service in France in the 1980s. For several months, this area has presented a number of track maintenance anomalies such as mud pumping in the ballast and geometry faults (longitudinal levelling and gauge). The mud pumping phenomenon leads to ballast fouling and accelerated deterioration of the track geometry. Due to repeated train loading, fine particles and water are driven upward from the subgrade into the ballast due to insufficient stiffness of RT layers and inadequate drainage. Despite regular maintenance interventions on this zone (tamping, local ballast renewal, geotextile installation) the track continues to exhibit significant issues.  

The section of track affected by this recurrent and abnormal maintenance extends over 130~m to the east of the zone. The position of this zone named ‘disorder zone’ in following, is represented on the Figure~\ref{fig:map}a and identified by Kilometre Points~(KP), the exact position of which is confidential. In this specific zone, all potential causes related to track components, such as defective rails or sleepers, were investigated, but no anomalies were found. Attention then shifted to the hypothesis of a mechanical issue involving the RT structure or the shallow RE. The persistent degradation of track geometry can then be attributed to the poor quality of the sub-ballast layer, capping layer and/or subgrade.  

% % [Figure 1]

\subsection*{Archive and conventional diagnostic data}
\begin{figure}
    \includegraphics[width=0.9\linewidth]{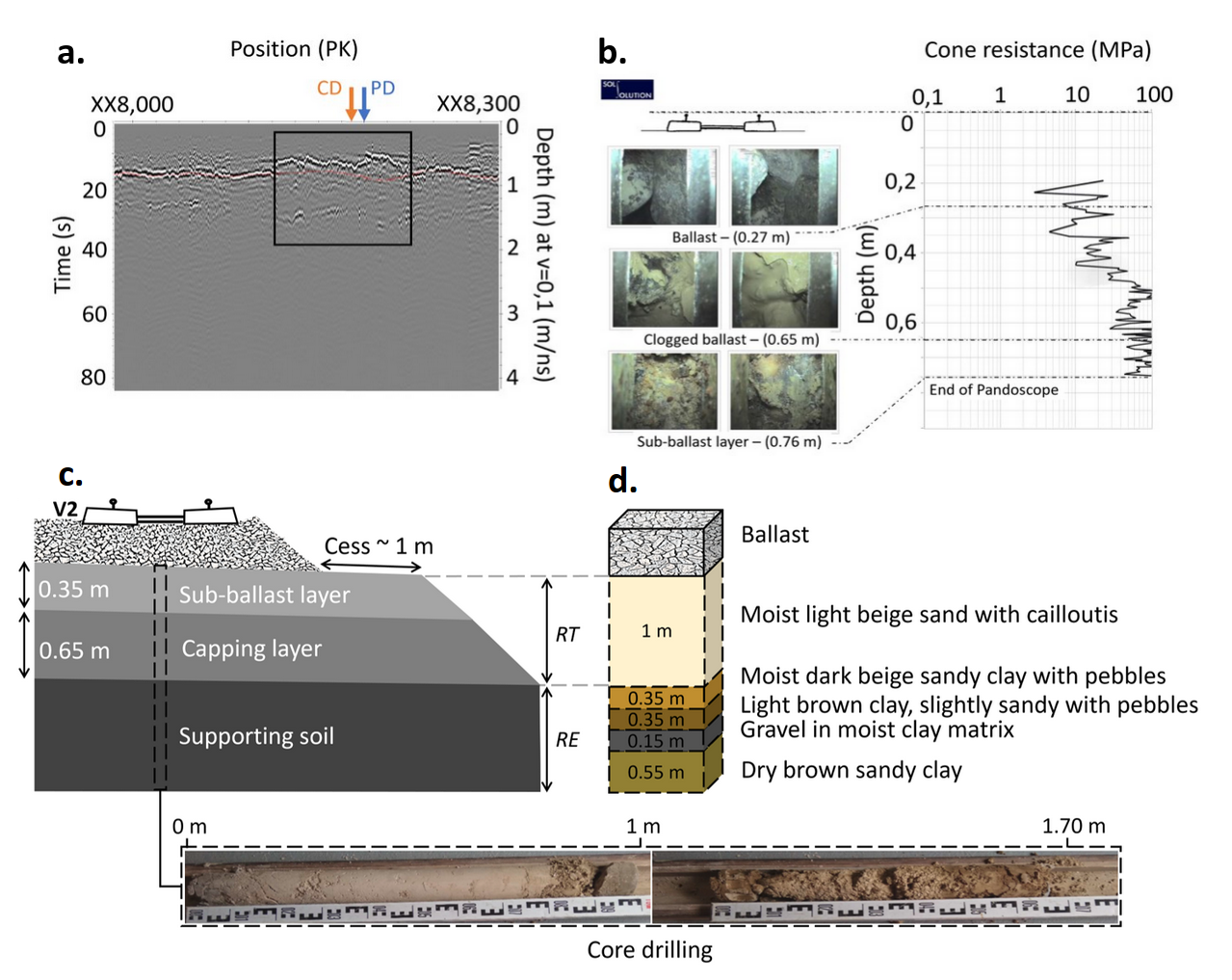}
    \centering
    \caption{\textbf{a.} Archive radargram of the extended study zone. The area framed in black corresponds to the disorder zone with the signature of mud pumping in the ballast. \textbf{b.}~Archive DCP with endoscopy technique survey with diagram of cone resistance as a function of depth and the photographs associated with each layer. \textbf{c.}~ Schematic interpretation of RE structure from archive data and photographies of soil cores from 0.8 (under the top of the ballast) to 1.8~m depth and from 1.8~m to 2.5~m. \textbf{d.} Interpretation of the entire soil core from the borehole.}
    \label{fig:archive_data}
\end{figure}

To assess this, both geotechnical investigations (core drilling and DCP tests) and ground-penetrating radar (GPR) were conducted in the affected zone. Their respective positions are shown in Figure~\ref{fig:map}a. Regular rounds of track surveillance (through visual monitoring and monthly auscultation by GPR) have made it possible to determine a singular zone where track disorders are appearing (Figure~\ref{fig:archive_data}a). The GPR monitoring of the HSL showed that there was an anomaly in the RT (ballast, sub-ballast layer and capping layer) with a high level of humidity and the signature of mud. This GPR zoning made it possible to place two types of survey in the area particularly affected by the disorder: a DCP test with endoscopy technique and a core drilling with laboratory analysis. The results of the DCP survey, shown in Figure~\ref{fig:archive_data}b, confirmed the presence of a heavily clogged ballast layer with mud, which originated at depth. The cone of the DCP also penetrated the sub-ballast layer at 0.65~m depth. According to the dimensioning standards, the compaction of the sub-ballast layer should be such that the penetrometer should not be able to penetrate it (as it is manually pushed). This results indicate weak mechanical properties of the sub-ballast layer in this zone. 

According to SNCF Réseau internal references and archive data (building standards of the construction time), the RT consists of a 0.35~m sub-ballast layer and a 0.65~m capping layer, all on top of the existing subgrade (Figure~\ref{fig:archive_data}c). Analysis of the core has made it possible to interpret the structure of the RT and identify the type of supporting soil linked to the local geology. The RE geology consist of beige sand over alternating sandy clays (Figure~\ref{fig:archive_data}d). 

% % [Figure 2]

\subsection*{MASW deployment}
In the context of RE diagnostics, the objective is to evaluate potential variations in $V_s$ profiles and to investigate whether a specific $V_s$ profile structure could account for the measured track geometry degradation. To this end, the MASW acquisition strategy employed for characterizing the maintenance anomaly zone relies on spatial profiling. This approach focuses on identifying changes in the $V_s$ parameter or detecting anomalies in the vertical $V_s$ distribution along the affected section, such as the presence of a LVL. Two seismic arrays have then been placed in relevant positions: on a zone not affected by the the track geometry anomaly, named P0 (west in Figure~\ref{fig:map}a) and the other on the disorder zone named P1 (east in Figure~\ref{fig:map}a). These measurements on the cess are based on the assumption that it is made of the same materials as the RT (Figure~\ref{fig:archive_data}c). Theses seismic arrays were first performed on the cess in order to guarantee good coupling between geophones and soil \citep{Gunn2016,Burzawa2023}. 

Each survey is made up of 96~vertical component geophones (14~Hz low cut-off frequency) spaced 0.25~m apart ($\Delta$x), which formed a 23.75~m long setup. The seismic shots were performed with an impact of 1.25~kg sledgehammer on a metal plate. Each seismic shot corresponded to the stack of 3~impacts of sledgehammer/plate to improve the signal-to-noise (S/N) ratio. The seismic shots were conducted at both ends of the array, referred to as the ‘direct shot’ (at the near end) and the ‘reverse shot’ (at the far end) with an offset of 0.125~m, corresponding to half the inter-geophone spacing. This slight offset was defined to minimise near-field effects and ensure clearer wavefield recordings at the first receivers \citep{Bodet2005}. The recording length was 2~s with a sampling interval of 0.5~ms (2000~Hz frequency sampling) and a pre-triggering delay of 0.02~s (Table~\ref{tab:param_classique_LS}).

Subsequently, the exact positions of P0 and P1 were used to deploy and evaluate the LS on the cess (Figure~\ref{fig:map}c). The LS setup actually consists of 48~geophones (4.5~Hz vertical geophones of lower frequency than for the conventional setup) spaced with 0.5~m mounted to steel plates and connected via an inelastic strap. The acquisition geometry was therefore similar in length to the conventional setup described above, but with half as many sensors and twice the receiver spacing. The sampling interval fixed at 0.125~ms (a sampling frequency of 8000~Hz) is 4 times greater than for the conventional setup (Table~\ref{tab:param_classique_LS}). 

The seismic shots for acquisition on the cess were performed with a stack of 3 impacts of hammer on a large polyester plate. However, for track acquisition with the LS, the seismic source was an automated weight drop source (13.7~kg, 0.965 m drop height), hitting a nylon strike plate base with a stack of 2 shots. The strategy of roll-along, involves progressively shifting the LS along the track by 6~m and performed direct shot only at each position with a 9~m offset was used on the track (Figure~\ref{fig:map}d and Table~\ref{tab:param_classique_LS}). The LS was towed behind a rail track trolley. The seismic signal recording time for the LS is defined as 1.5~s with a pre-triggering delay of~-0.01~s. The acquisition parameters for the conventional setup were defined \textit{prior} to the comparative study, based on a preliminary investigation aimed at optimizing data quality for the given context \citep{Burzawa2023}. In contrast, the acquisition parameters for the LS were primarily constrained by the available equipment and adapted to the specific constraints of the track itself, where objects such as concrete slabs may be encountered. Therefore, a higher sampling frequency is necessary to properly capture and identify these features.

% [Tableau 1 : paramètres d'acquisitions sismiques]
\begin{table}[]
\label{tab:param_classique_LS}
\caption{List of acquisition parameters for conventional seismic setup and landstreamer.}
\resizebox{\textwidth}{!}{%
\begin{tabular}{lccc}
\hline
\textbf{Parameters}                 & \textbf{Conventional setup (cess)}    & \textbf{Landstreamer (cess)} & \textbf{Landstreamer (track)}\\ \hline
Number of geophone                  & 96                                    & 48                           & 48               \\
Geophone low cut-off frequency (Hz) & 14                                    & 4.5                          & 4.5              \\
Receiver spacing $\Delta$x (m)                       & 0.25                                  & 0.5                          & 0.5              \\
Length (m)                          & 23.75                                 & 23.5                         & 23.5             \\
Sampling interval (ms)              & 0.5                                   & 0.125                        & 0.125            \\
Frequency sampling (Hz)             & 2000                                  & 8000                         & 8000             \\
Recording time (s)                  & 2                                     & 1.5                          & 1.5              \\
Delay (s)                           & -0.02                                 & -0.01                        & -0.01            \\ 
Source type                         & Sledgehammer                          & Hammer                       & Weight drop       \\
Coupling plate                      & Metal plate                           & Nylon strike plate           & Nylon strike plate  \\
Position of seismic shot            & Direct/reverse                        & Direct/reverse               & Direct             \\
Stack                               & 3                                     & 3                            & 2                   \\
Offset (m)                          & $\Delta$x/2                           & $\Delta$x/2                  & 9                  \\
Roll-along setup shift (m)                      & --                                    & --                           & 6                   \\  \hline
\end{tabular}%
}
\end{table}

\section*{Data variability along the earthwork}
\label{sec:inversly_dispersive_media}
\subsection*{Comparison of raw data}
The vertical particle displacement velocities generated at the ground surface, in response to controlled seismic sho, was recorded in the distance-time domain ($d - t$) and then represented in figures called seismograms (or shot gathers). 
 
\begin{figure}
    \includegraphics[width=1\linewidth]{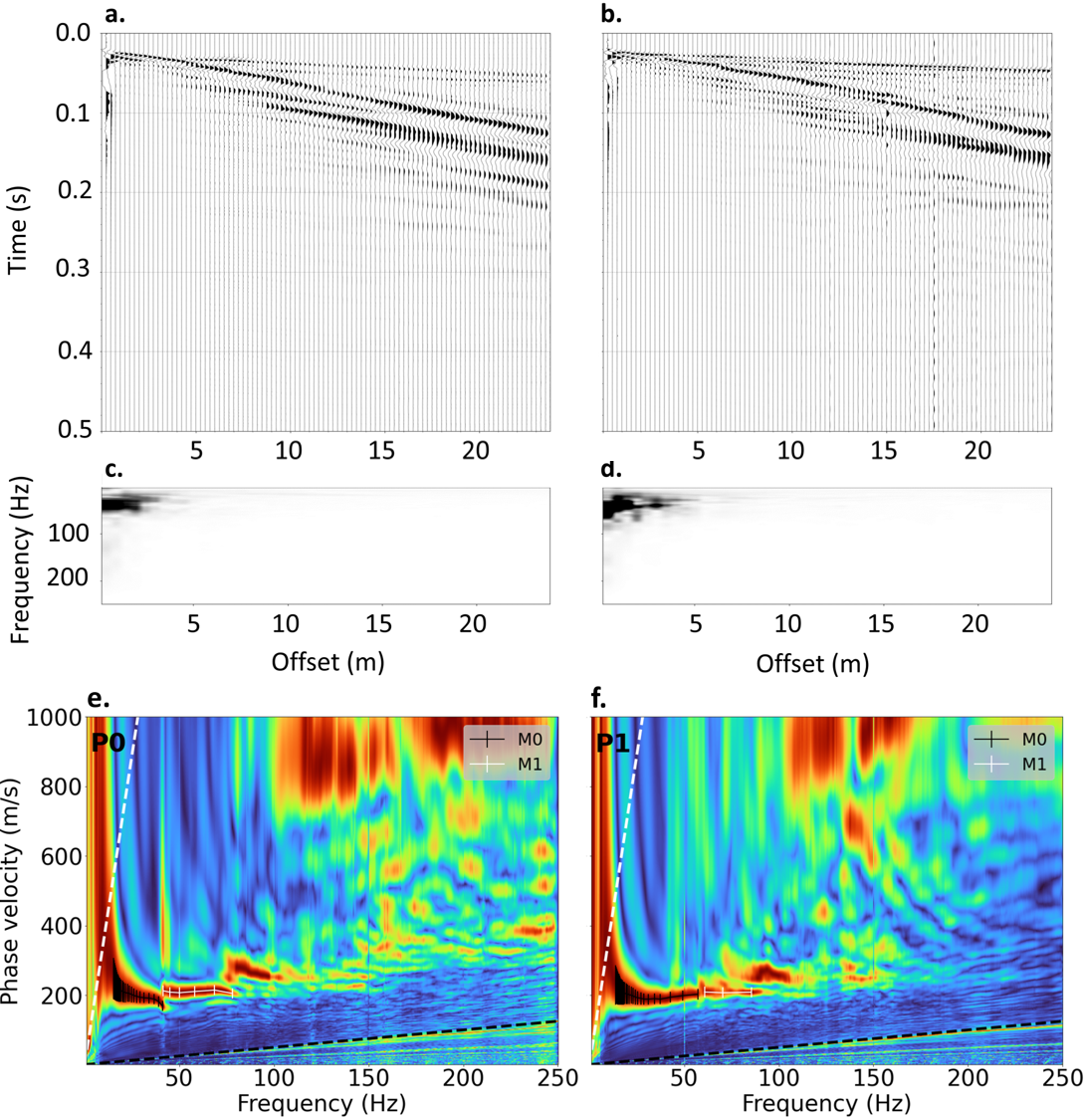}
    \centering
    \caption{\textbf{a.} and \textbf{b.} Seismograms acquired with the conventional setup on the cess of the direct shot at position P0 and P1, respectively. Associated spectrogram for P0 (\textbf{c.}) and P1 (\textbf{d.}). Dispersion images for P0 (\textbf{e.}) and P1 (\textbf{f.}). Dispersion images of the direct and reverse shots were stacked. The fundamental mode (M0) is shown in black and the first higher mode (M1) in white. Limits of dispersion image analysis ($\lambda_{min}$ and $\lambda_{max}$) are plotted in black and white dashed lines respectively. The picked dispersion curves were resampled in wavelength and the error calculated according to the relation of \cite{ONeill2003}.
    }
    \label{fig:P0-P1_cess_conventionnal}
\end{figure}

For the two positions (P0 and P1), the seismograms and corresponding frequency content represented in spectrograms, are presented in Figure~\ref{fig:P0-P1_cess_conventionnal}a to d. A comparison of the direct shot from both positions reveals that the surface-wave field, clearly identifiable in both cases, exhibits similar apparent velocities. However, the frequency content is notably lower at P0. The P-wave train can also be distinguished, particularly at P1. Both seismograms exhibit noise as well as a few low-quality traces, likely due to poor local coupling between some geophones and the ground. The spectrograms confirm a concentration of high-energy content between 20 and 50~Hz in the vicinity of the seismic source for both positions.

% % [Figure 3]
% %

\subsection*{Differences between extracted information along the cess}
The dispersive nature of surface waves is analysed in the frequency-phase velocity domain ($f - V_{\phi}$), using the slant-stack wavefield transposition method \citep{Mokhtar1988}. The result of such transform is called a ‘dispersion image’ in the following. Several geophone window sizes were tested to assess lateral variations and ensure the validity of the 1D assumption required for inversion. Based on the comparison of dispersion images from direct and reverse shots, a maximum window size of 96 geophones was selected. This configuration showed minimal sensitivity to lateral heterogeneities, ensuring sufficient spatial coherence for reliable analysis \citep{Bodet2005,Steinel2014,Burzawa2023}. Given the limited lateral variability, stacking dispersion images from both shots was performed to enhance the S/N ratio \citep{Pasquet2017,Burzawa2023}. The analysis limits of the dispersion images are defined by the minimum and maximum identifiable wavelengths ($\lambda_{min}$ and $\lambda_{max}$, respectively). $\lambda_{min}$, is governed by the geophone spacing and is commonly set as $2\Delta x$. It must be sufficiently small to prevent spatial aliasing and to ensure reliable resolution of high-frequency content. $\lambda_{max}$, is empirically defined as approximately the setup length.

To facilitate analysis, dispersion images were normalised at each frequency. The dispersion images shown in Figure~\ref{fig:P0-P1_cess_conventionnal}e and f illustrate very good resolution of the surface-wave dispersion at both positions. The energy maxima correspond to the fundamental mode (M0) and the first higher mode (M1) of Rayleigh-wave propagation. These curves, named ‘dispersion curves’, were manually and subsequently resampled in terms of~$\lambda$. The associated uncertainties, represented as error bars, follow a Lorentzian distribution at low frequencies, as suggested by \citet{ONeill2003a}. At position P0, M0 shows a decrease of $V_{\phi}$ with an increasing of $f$, whereas at position P1, M0 shows a significant increase at 50~Hz. M1 is comparable to the two positions.

\section*{Improving acquisition efficiency with the landstreamer}
\label{sec:landstreamer}
\begin{figure}
    \includegraphics[width=1\linewidth]{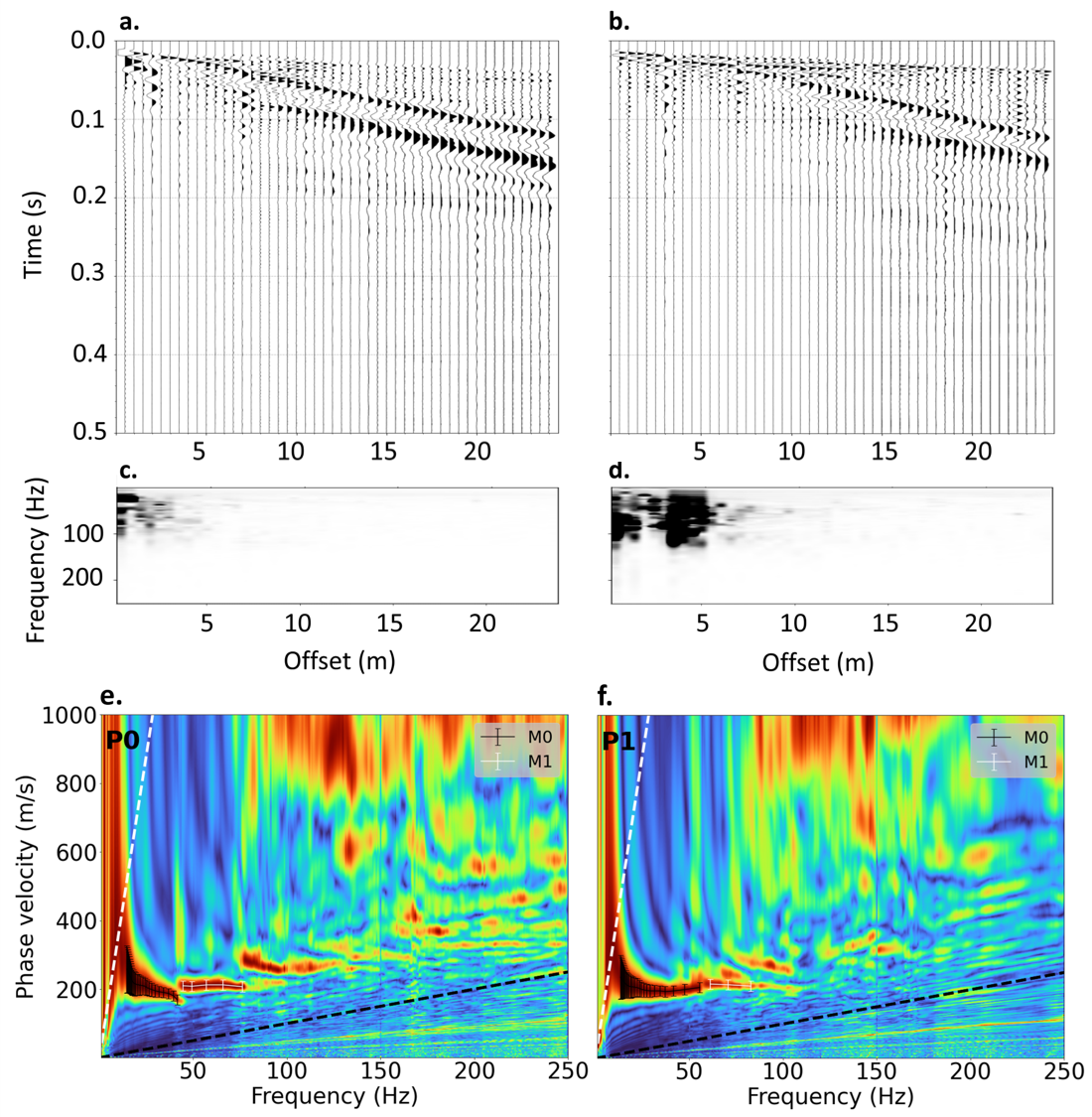}
    \centering
    \caption{\textbf{a.} and \textbf{b.} Seismograms acquired with the Landstreamer setup on the cess of the direct shot at position P0 and P1, respectively. Associated spectrogram for P0 (\textbf{c.}) and P1 (\textbf{d.}). Dispersion images for P0 (\textbf{e.}) and P1 (\textbf{f.}). Dispersion images of the direct and reverse shots were stacked. The fundamental mode (M0) is shown in black and the first higher mode (M1) in white. Limits of dispersion image analysis ($\lambda_{min}$ and $\lambda_{max}$) are plotted in black and white dashed lines respectively. The picked dispersion curves were resampled in wavelength and the error calculated according to the relation of \cite{ONeill2003}.
    }
    \label{fig:P0_P1_cess_landstreamer}
\end{figure}

In order to compare the two acquisition setups, data measured with the LS were acquired at the same position as P0 and P1 on the cess. The seismograms are shown in Figure~\ref{fig:P0_P1_cess_landstreamer}a and b for positions P0 and P1 respectively. At both locations, P waves and surface waves are identifiable. Few traces are saturated or significantly degraded. Surface waves recorded at P1 appear noticeably more attenuated compared to those at P0. Figure~\ref{fig:P0_P1_cess_landstreamer}c and d show the corresponding spectrograms. The spectrogram for the direct shot at position P1 reveals a very strong amplitude at short offsets (less than 5~m), a phenomenon that is not observed at position P0. The dispersion images resulting from the stacking of direct and reverse shots are shown in Figure~\ref{fig:P0_P1_cess_landstreamer}e and f. These images show good resolution of surface-wave dispersion and are used to extract dispersion curves (M0 and M1). In the same way as above with the conventional setup, M0 displays decreasing smoothly with frequency at P0. At P1, however, M0 exhibits an increases with frequency. 

% % [Figure 4]
% %

\subsection*{Comparison with the conventional setup on the cess}
\label{subsec:landstreamer_cess}
\begin{figure}
    \includegraphics[width=1\linewidth]{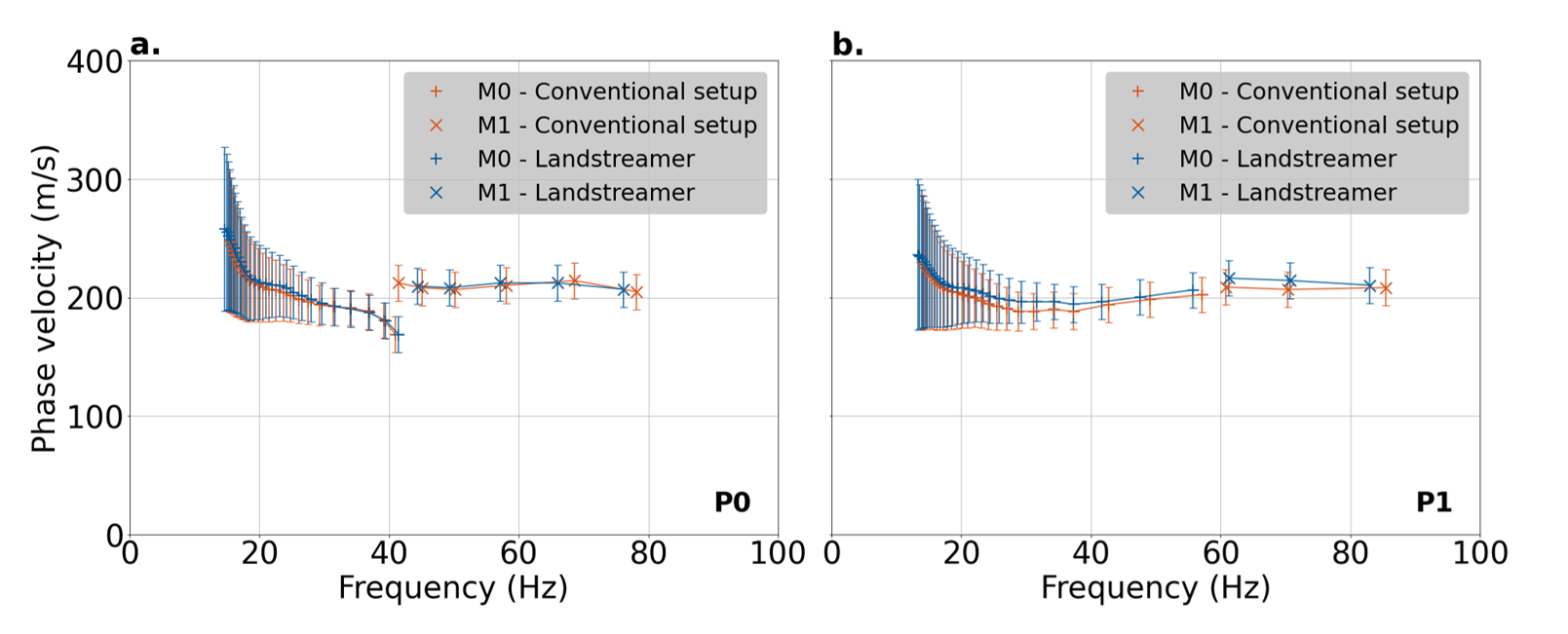}
    \centering
    \caption{Picked dispersion curves from the conventional setup (orange) and from landstreamer (blue) for P0 and P1 respectively. The fundamental mode (M0) is represented by~‘+’ symbol and first higher mode (M1) by~‘x’ symbol. The picked dispersion curves were resampled in wavelength and the error bars were calculated according to a Lorentzian distribution type at low frequencies \citep{ONeill2003a}.
    }
    \label{fig:comparison_conventional-LS}
\end{figure}

To make a detailed comparison of the two seismic setups, the seismograms of the conventional setup was degraded with the selection of one geophones out of two, a cut-off at 1.5~s and down sampled to~2000~Hz. As our research is not conducted within the $d - t$ domain, a detailed analysis of data in this domain has not been undertaken. The seismograms presented above were transposed into $f - V_{\phi}$ domain and dispersion curves of M0 and M1 were identified. Figure~\ref{fig:comparison_conventional-LS}a and b represents the extracted dispersion curves for the two setups at position P0 and P1, respectively. The comparison of these curves (M0 and M1) reveals a high degree of similarity. The differences observed lie within the estimated uncertainties \citep{ONeill2003}. 

% % [Figure 5]
% %

%
\subsection*{Deployment of the landstreamer on the track}
\label{subsec:landstreamer_track}
After confirming that the towed seismic setup provided good quality data for characterising the RE, the LS was deployed on the track (i.e., on the ballast). Indeed, the dynamic and mechanical stresses on the track are much greater than on the cess due to the number of passing trains and their speeds, particularly on HSL. The acquisition was carried out at night when traffic was interrupted. A roll-along acquisition strategy was implemented, as already detailed in MASW deployment section, enabling the area to be covered over 300~m. 

\begin{figure}
    \includegraphics[width=1\linewidth]{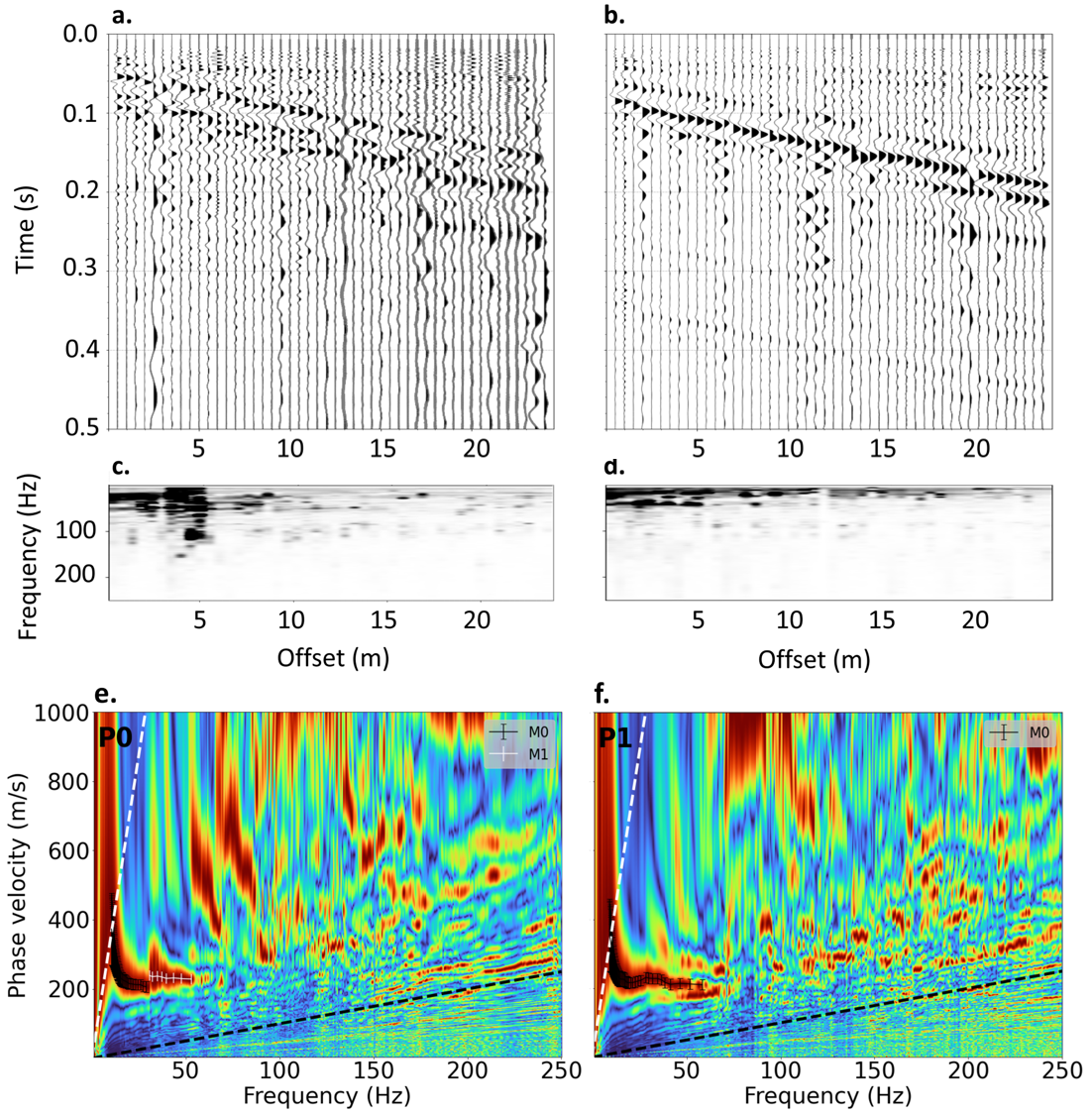}
    \centering
    \caption{Seismogram acquired with the landstreamer on the track at the position P0 (\textbf{a.}) and P1 (\textbf{b.}). Associated spectrogram of P0 (\textbf{c.}) and P1 (\textbf{d.}). Dispersion images of seismic survey at the position P0 (\textbf{e.}) and P1 (\textbf{f.}). The direct and reverse shots were stacked. The fundamental mode (M0) is shown in black and the first higher mode (M1) in white. Limits of dispersion image analysis ($\lambda_{min}$ and $\lambda_{max}$) are plotted in black and white dotted lines respectively. The picked dispersion curves were resampled in wavelength and the error calculated according to the relation of \cite{ONeill2003}.
    }
    \label{fig:track_landstreamer}
\end{figure}

The data presented in this section were selected at track equivalent positions of P0 and P1. The seismograms, associated spectrograms and dispersion images for these two zones are shown in Figure~\ref{fig:track_landstreamer}. The seismograms show a higher level of noise, especially so for those located in the P0 zone (Figure~\ref{fig:track_landstreamer}a). The seismograms of the P0 zone reveals numerous traces characterized by elevated levels of noise and degraded signal (traces n°5 and n°96). The P-wave train are also noisy in both zone. The seismograms of the P1 zone (Figure~\ref{fig:track_landstreamer}b) show probable source bouncing effect at around 0.3~s. The dispersion images corresponding to the single performed shot (Figure~\ref{fig:track_landstreamer}e and f) are, in the same way as the seismograms, particularly noisy compared to the acquisitions performed on the cess. The dispersion curves are identifiable up to 50~Hz, but beyond that, the dispersion impossible to picked precisely. M0 is identifiable in both dispersion images, but M1 is only picked in the dispersion image from P0 zone. 

% % [Figure 6]
% %

\begin{figure}
    \includegraphics[width=1\linewidth]{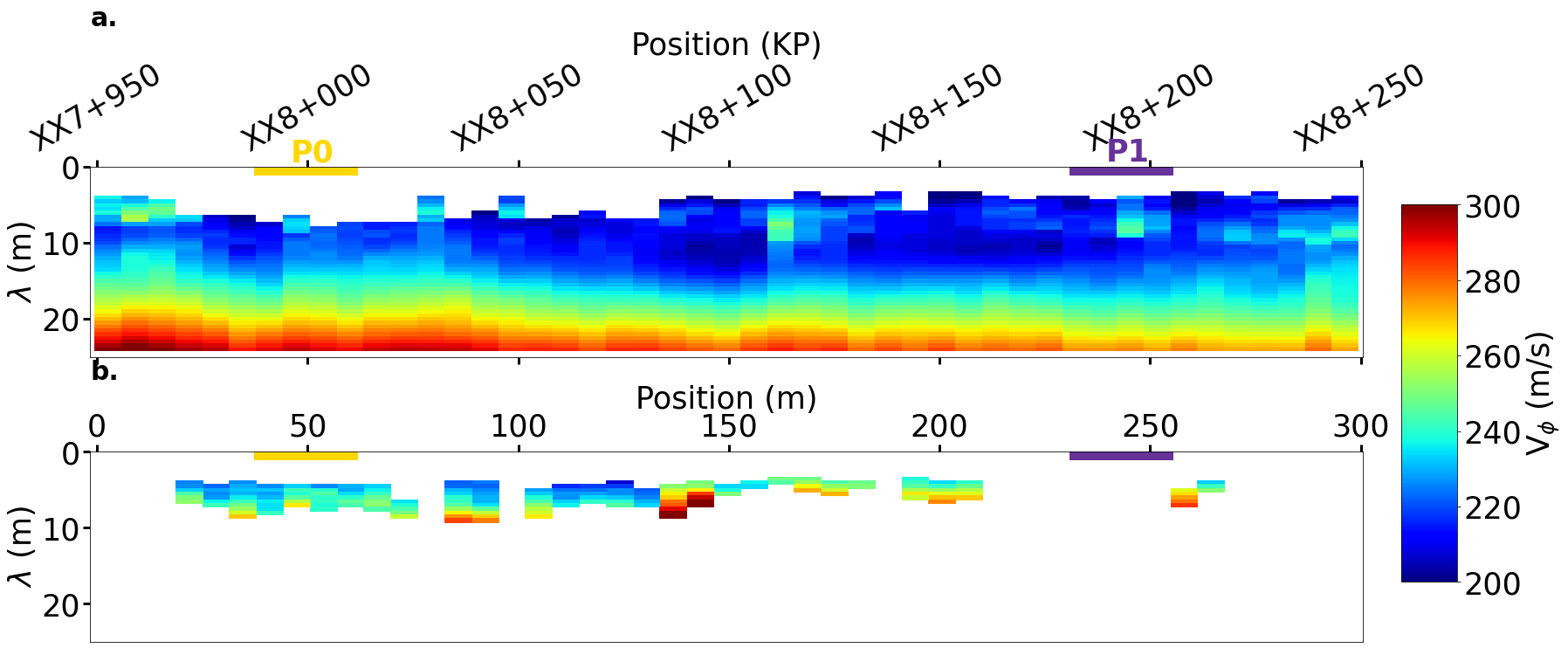}
    \centering
    \caption{\textbf{a.} Picked dispersion curves for the fundamental mode (M0) plotted as a pseudosection as a function of position along the track and wavelength ($\lambda$). \textbf{b.} Picked dispersion curves for the first higher mode (M1) plotted as a pseudosection as a function of extraction position along the track and $\lambda$. The $\lambda_{max}$ associated with the average value of the minimum frequency identified for M0 along the section is estimated at 25~m.
    }
    \label{fig:vphi_M0_M1}
\end{figure}

This step of quality control of the seismograms and dispersion images was carried out for all the 47 dispersion images acquired during the roll-along on the track. All 47 dispersion images were picked for M0 in the range from 10 to 40~Hz minimum. An accurate identification of M1 is essential for improving the resolution of near-surface characterization. M1 was extracted only in images where it could be clearly identified. The dispersion curves for each step of the roll-along were plotted in the $\lambda$-$V_{\phi}$ domain, with $V_{\phi}$ associate to a colour, and juxtaposed in acquisition order to form a pseudo-2D $V_{\phi}$ section. $\lambda_{max}$ associated with the length of the setup is here estimated at 25~m. M0 and M1 pseudo-2D $V_{\phi}$ sections are represented, respectively in Figure~\ref{fig:vphi_M0_M1}a and b. The presence of M1 was identifiable only in certain dispersion images, particularly those acquired between positions 25 and 210~m (Figure~\ref{fig:vphi_M0_M1}b). Given the uniform flat topography across the study area, no elevation corrections were applied to the section. Each 1D $V_{\phi}$ profile begins at $\lambda$ corresponding to the maximum of the high-frequency pick of dispersion curves, which represents the shallowest investigated layer.

% % [Figure 7 : Pseudo section of phase velocity for M0 and M1]
% %

\subsection*{Bayesian inversions results}

The pseudo-2D $V_s$ section is constructed by juxtaposing 1D surface-wave inversions performed at each location along the profile, using the local dispersion curves extracted from the pseudo-2D $V_{\phi}$ section. Relying on a single ‘best-fitting’ model can be misleading, particularly when the result is influenced by \textit{prior} regularization choices or incorrect noise assumptions \citep{Sambridge2013}. To overcome these limitations, this study adopts the Bayesian inversion framework. The aim of Bayesian inference is to quantify the \textit{posterior} distribution which is the probability density of the model parameters \citep{Bodin2012}. The algorithm used in this study was developed by \cite{Magrini_etal2025} and employs a Reversible Jump Markov chain Monte Carlo (RJ-McMC) \citep{Green1995,Green2003} to explore the model space with a uniform \textit{prior} probability density.

The parametrization is defined by the number of layers and each $H$, pressure-wave velocity ($V_p$), $V_s$ and density $\rho$. In the current study, it was defined according to \textit{prior} from the archives and the input data (from the geotechnical and geophysical surveys described earlier as well as typical characteristics of the RT). The RT and shallow RE structure is made up of 4 layers over an infinite half-space: ballast layer, sub-ballast layer, capping layer and subgrade. $H$ of the ballast layer and sub-ballast layer ($H_{1}$ and $H_{2}$ respectively) can vary between 0.5 and 3~m, and that of the capping layer up to 5~m. The $V_s$ of each layer can vary from 80~m/s to 500~m/s, except for half-space, whose $V_{s4}$ can vary from 100 to 600~m/s. The associated $V_p$ is computed with a $\sfrac{V_p}{V_s} = 2$ that correspond to Poisson coefficient $\nu =0.33$. The density of all layers is fixed at 2000~kg/m$^3$. The standard deviations (\textit{std}) of the Gaussian proposal functions for each parameters are fixed at 5~m/s for $V_s$ and 0.05~m for $H$. \textit{std} was fixed after trial-and-error to have an acceptance rate close to 44$\%$ \citep{rosenthal2000}. Once this parameter space defined, the RJ-MCMC algorithm is launched with: 5 chains of 150 000 iterations and 30 000 iterations for the burn-in period. 

\begin{figure}
    \includegraphics[width=0.8\linewidth]{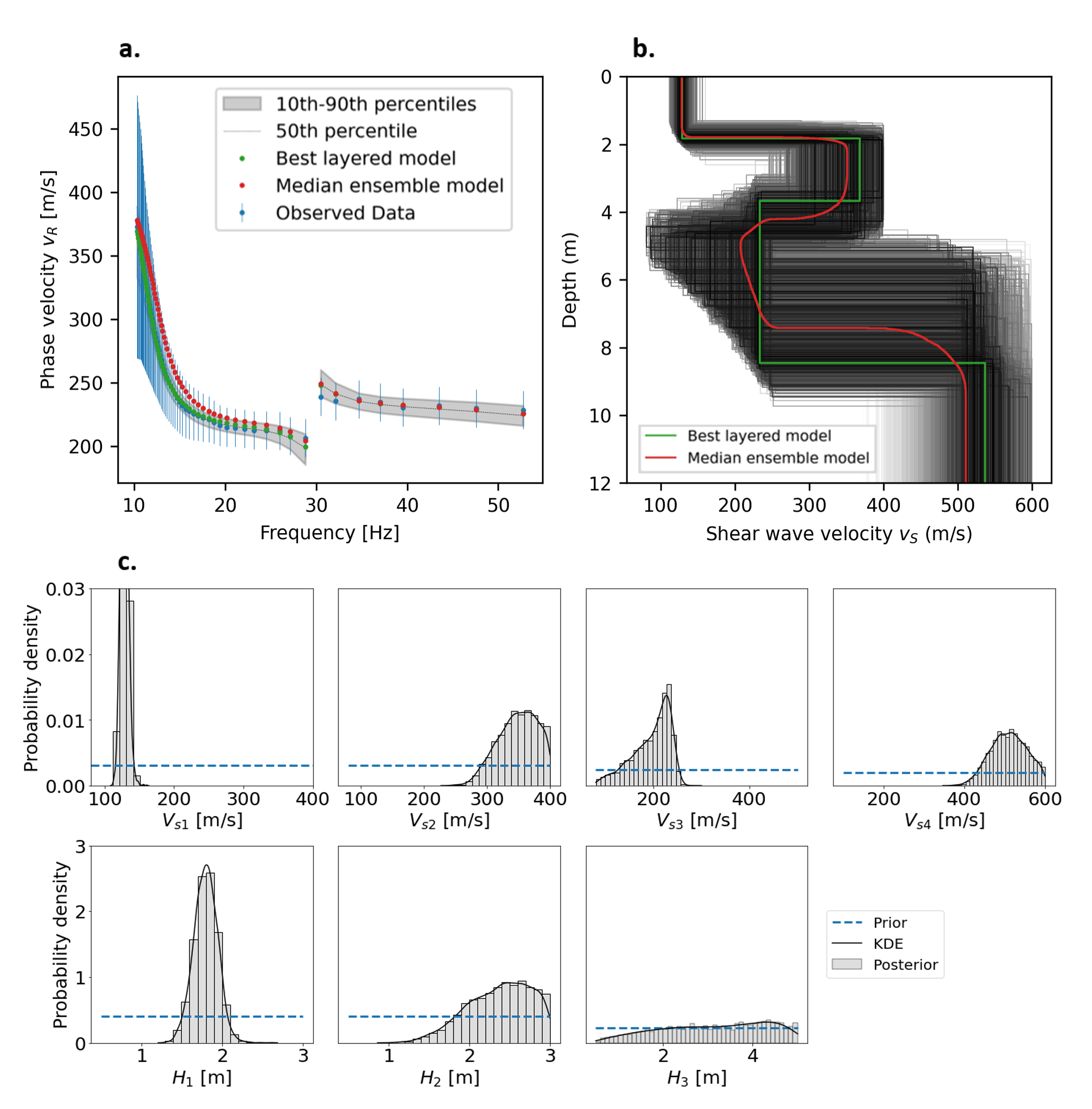}
    \centering
    \caption{Results of the inversion at position P0 on the track. \textbf{a.} Observed (blue) and calculated dispersion curves. The dispersion curve associated with the median of all the models explored is represented in red and the dispersion curve associated with the best layered model is represented in green. The grey area represents the 10-90th percentiles of the models explored. \textbf{b.} $V_s$ The 1D $V_s$ profiles calculated during inversion are shown in black. The median model is shown in red and the best layered model in green. \textbf{c.} Histograms for each parameter of the inverse problem ($V_{s1}$, $V_{s2}$, $V_{s3}$, $V_{s4}$, $H_{1}$, $H_{2}$ and $H_{3}$). The bin width is equal to twice the \textit{std}. The blue dashed line represents the uniform marginal \textit{prior} pdf and the black line represents the KDE of each distribution.
    }
    \label{fig:inversion_1D_P0}
\end{figure}

For position P0, the observed dispersion curve is shown in blue with associated error bars (Figure~\ref{fig:inversion_1D_P0}a). The shaded grey area represents the 10th–90th percentile envelope of the calculated dispersion curves. The red markers corresponds to the median model from the ensemble, while the green markers indicate the best-fit layered model. A good agreement is observed between the measured data and the predicted models. 1D $V_s$ profiles are plotted as a function of depth in Figure~\ref{fig:inversion_1D_P0}b. The ensemble of acceptable models is shown in black lines, highlighting the variability across solutions. The median profile (red) and the best-fit layered model (green) are also displayed for comparison. The 1D  $V_s$ profiles show well-constrained structures at shallow depth, with $V_{s1}$ values converging toward 120~m/s. The second layer exhibits $V_{s2}$ exceeding 350 m/s, followed by a third layer, $V_{s3}$, around 230~m/s, and finally a half-space with $V_{s3}$ approaching 500~m/s. \textit{Posterior} distributions for each parameter ($V_{s1}$ to $V_{s4}$ and $H_{1}$ to $H_{3}$) are presented as histograms with associated kernel density estimates (KDE). The histograms are computed using a bin width equal to twice the \textit{std} of the corresponding distribution. The KDE is used to estimate the underlying probability density function (pdf) \citep{parzen1962estimation}. Consistent with the previously described 1D $V_s$ profiles, $V_s$ \textit{posterior} pdf for all layers show good resolution of the parameter. The parameter $H_{1}$ is well resolved, with the \textit{posterior} indicating a thickness of approximately 1.8~m. The parameter $H_{2}$ is estimated at 2.8 m, while $H_{3}$ is not well constrained on the explored space.

% % [Figure 8]
% %

\begin{figure}
    \includegraphics[width=0.8\linewidth]{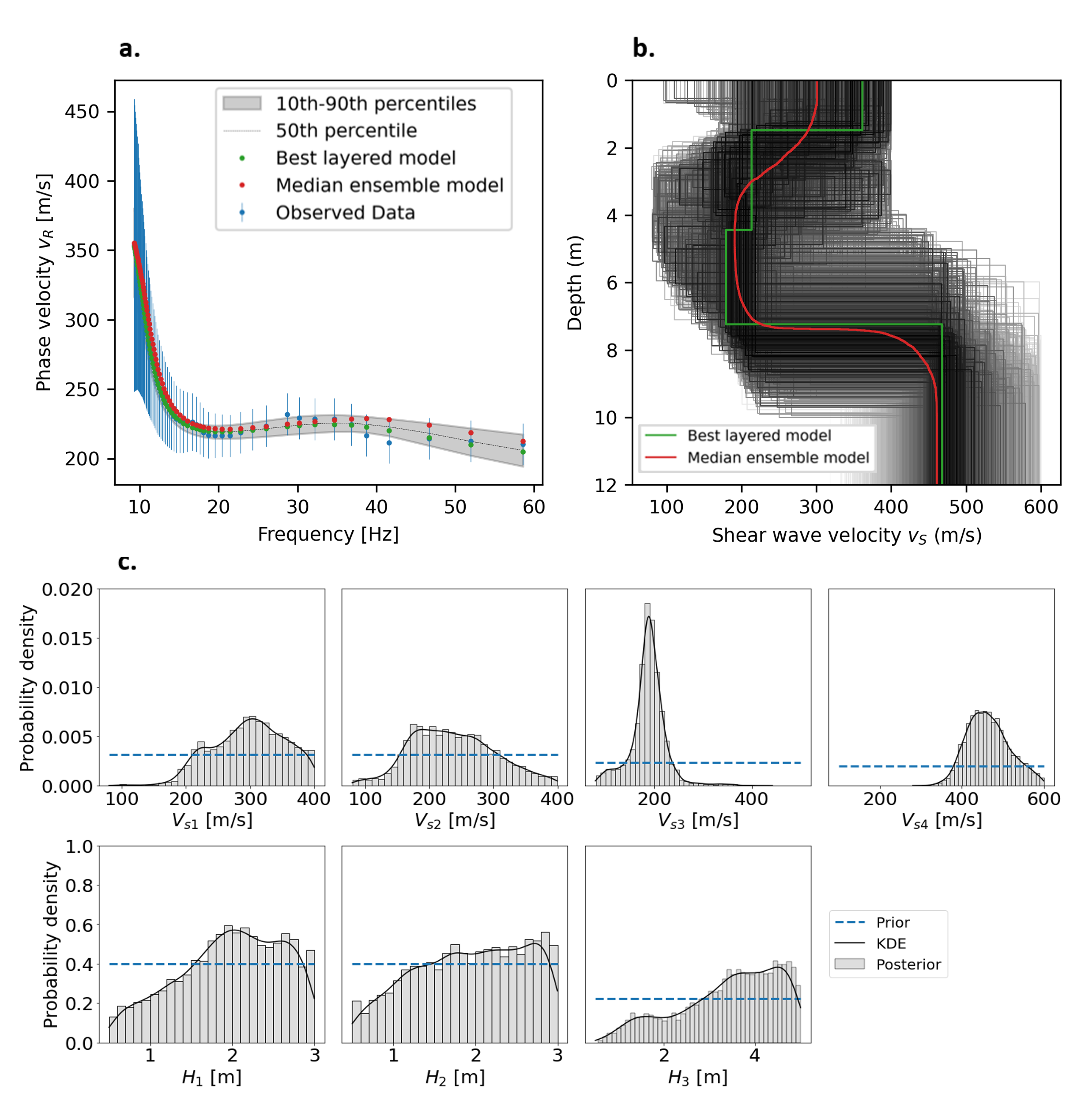}
    \centering
    \caption{Results of the inversion at position P1 on the track. \textbf{a.} Observed (blue) and calculated dispersion curves. The dispersion curve associated with the median of all the models explored is represented in red and the dispersion curve associated with the best layered model is represented in green. The grey area represents the 10-90th percentiles of the models explored. \textbf{b.} $V_s$ The 1D $V_s$ profiles calculated during inversion are shown in black. The median model is shown in red and the best layered model in green. \textbf{c.} Histograms for each parameter of the inverse problem ($V_{s1}$, $V_{s2}$, $V_{s3}$, $V_{s4}$, $H_{1}$, $H_{2}$ and $H_{3}$). The bin width is equal to twice the \textit{std}. The blue dashed line represents the uniform marginal \textit{prior} pdf and the black line represents the KDE of each distribution.
    }
    \label{fig:inversion_1D_P1}
\end{figure}

For position P1, calculated dispersion curves compared to the observed data are represented in Figure~\ref{fig:inversion_1D_P1}a. The 10th–90th percentile envelope remains narrow across the entire frequency range, indicating strong convergence among the sampled models. The median model (red) and the best-fit layered model (green) closely follow the observed trend. The 1D $V_s$ profiles show an equivalent solution density over the entire investigated depth. A velocity inversion is noticeable at a depth of 2~m and a significant thickness of 5~m. The median model exhibits a gradual velocity decrease, while the best-fit model highlights distinct velocity contrasts at 2~m. The \textit{posterior} distributions of the inverted parameters display varied behaviours between each parameters. $V_{s1}$ and $V_{s2}$ show broad distributions, indicating greater uncertainty, while $V_{s3}$ is sharply peaked around 200~m/s, suggesting that this layer is well resolved. The layer thicknesses ($H_{1}$ to $H_{3}$) show spread distributions, demonstrating the low resolution of this parameter.

% % [Figure 9]

\begin{figure}
    \includegraphics[width=1\linewidth]{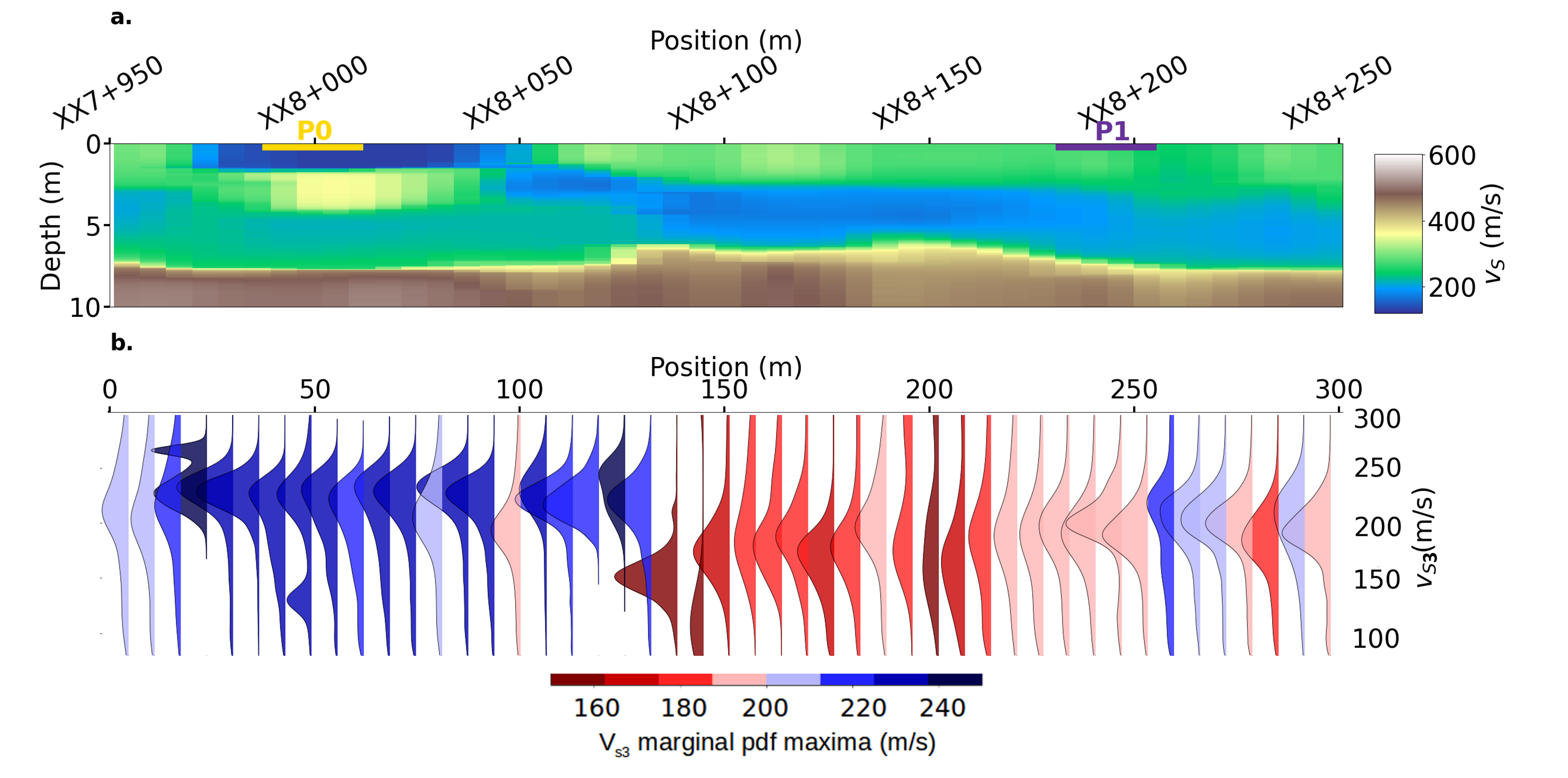}
    \centering
    \caption{\textbf{a.} Pseudo-2D $V_s$ section constructed by juxtaposing multiple 1D median models. Horizontal smoothing was applied using successive median filters with several window sizes (4, 3, and 2 adjacent points). The position of profiles P0 and P1 are shown in yellow and purple respectively.  
    \textbf{b.} Marginal 1D of the parameter $Vs_3$ along the line. The colour scale is represented according to the maximum estimated $Vs_3$ for each 1D inversion.
    }
    \label{fig:inversion_bayesbay}
\end{figure}

Each of the resulting 1D $V_s$ vertical profiles is extracted from the median ensemble. The pseudo-2D $V_s$ section is built by juxtaposing the 1D $V_s$ profile with lateral smoothing. Figure~\ref{fig:inversion_bayesbay} illustrates the spatial variability of the $V_s$ along the line. $V_p$ and $\rho$ are not considered in the following because of the poor influence on dispersion \citep{Pasquet2017}. The maximum depth of investigation ($Z_{max}$) associated with linear geometry is based on the empirical relationship of $\lambda_{max}$ divided by two or three \citep{Bodet2009,Pasquet2017,Foti2018}. A safe limit of $Z_{max}$ is set in this study as 10~m. 

Figure~\ref{fig:inversion_bayesbay}a presents the pseudo-2D $V_s$ section. The subsurface structure shows lateral heterogeneity, particularly in the upper 6~m, where $V_s$ ranges between~200 and 400~m/s. A LVL is particularly noticeable from position XX8+080 to XX8+250 with a $V_s$ of less than 200~m/s at depth of 2.5~m. In this zone, there is a substantial increase in the thickness of this layer to 3~m. The two investigated positions, P0 and P1, are marked along the profile, showing their location relative to this heterogeneous zone. Figure~\ref{fig:inversion_bayesbay}b shows the well-resolved \textit{posterior} distributions, namely that of the parameter $V_{s3}$ at each lateral position. The 1D marginal \textit{pdf} reveal localized velocity changes and distribution shapes that vary along the investigated zone. From 0 to 180~m, the \textit{pdf} shows average velocity ranging from 200 to 250~m/s. Some bimodal solutions appear but are weaker in amplitude. From 180 to 220~m, $V_{s3}$ decreases significantly, with values below 200~m/s. Additionally, the distributions widen and shift, indicating greater uncertainty. The colour code reinforces these trends, transitioning from blue (higher velocities) to red (lower velocities). 

% [Figure 10]

\section*{Discussions}
\label{sec:discussions}

\subsection*{LVL detection along the track}
As mentioned before, surface waves are sensitive to LVL \citep{Ryden2004}. In this context, the extracted dispersion curves often exhibit increased energy in higher propagation modes \citep{0neill_Matsuoka2005}. The extracted dispersion curves are typically composed of portions of higher modes rather than representing a single, continuous M0 \citep{Tokimatsu1992, Forbriger2003a, Forbriger2003b, Ryden2004}. This characterization introduces ambiguity and can lead to misinterpretation when identifying modal contributions during the picking process \citep{0neill_Matsuoka2005}. The presence of such LVL structure is particularly evident in HSL context, where the sub-ballast layer supporting the superstructure (rail-sleeper-ballast) is heavily compacted and rests on mechanically weaker underlying materials. 

The comparison between P0 and P1 acquired on the cess shows that although the overall $V_s$ ranges of M0 are similar in the two distinct zones, the shapes of the dispersion curves differ significantly. In particular, M0 at P1 shows a distinct decrease in $V_{\phi}$ with increasing frequency. This behaviour is characteristic of an inversely dispersive response to a medium. These significant differences in dispersion behaviour between P0 and P1 suggests localized differences in material stiffness or layering in RT. The contribution of higher propagation modes clearly improves the constraint on near-surface $V_s$. This is evidenced by the comparison between the 1D $V_s$ profiles at P0, where M1 was identified and included in the inversion, and those at P1, where only M0 was used. At P0, the near-surface structure is more sharply defined, with $V_{s1}$ values converging toward 120~m/s, indicating improved resolution in the shallow layers. In contrast, the model at P1 shows greater variability and uncertainty in the uppermost layers, highlighting the added value of incorporating higher modes when available.

The lateral evolution of $V_{\phi}$ reveals an increase at short $\lambda$ followed by a progressive decrease at longer $\lambda$, a pattern indicative of inverse dispersive behaviour. Inversion of the surface-wave dispersion on the track shows variations in this structure anomaly (decrease in $V_s$ and thickening of the layer). The notable evolution of this layer, characterized by its low $V_s$, is remarkably correlated with measurable track geometry degradation. The presence of this LVL issue is further substantiated by archive and conventional diagnostic data, which confirm mechanical degradation of the RT. Analysis of soil core samples reveals a relatively homogeneous layer approximately 1~m thick, likely resulting from the blending of sub-ballast and capping layers. This amalgamated layer fails to perform its intended mechanical roles.

\subsection*{Landstreamer benefits for trackbed mechanical properties access}
Despite the inherent differences between the conventional and LS setups, including variations in number of geophones and sampling frequencies, the results from both configurations show strong consistency. The extracted dispersion curves are remarkably similar and fall within the range of associated uncertainties. Minor discrepancies in the spectrograms are likely due to differences in seismic sources (sledgehammer vs. hand hammer). Nevertheless, with an appropriate processing, both seismic setups deliver high-quality and comparable dispersion data. 

The LS was then deployed directly onto the ballast layer, where seismic-wave propagation is influenced by significant energy attenuation within the ballast and complex interactions with the superstructure. A notable effect observed is the resonance of the sleepers along the zone when seismic shots are carried out in close proximity. Notable differences are also observed between dispersion curves acquired on the track and those from the cess. These differences are probably due to differences in subsurface conditions, influenced by repeated dynamic loads from high-speed train traffic. These loads and traffic modify the mechanical properties of shallow RT, resulting in a change in the behaviour of the dispersion. 

From a maintenance planning perspective, MASW data obtained using LS enable a shift from reactive to proactive maintenance strategies. By providing spatially continuous and physically grounded information on subsurface stiffness, the technique allows for early detection of degradation zones, enabling targeted and timely interventions. MASW surveys, particularly when conducted efficiently with LS systems, require minimal track occupation and offer a favourable cost-benefit ratio. The deployment of the LS setup for MASW acquisition presents significant operational and economic advantages in the context of railway maintenance. One of the most compelling benefits is the considerable reduction in track occupation time. For instance, surveying the 300~m section required approximately 10~h with the conventional setup, whereas the LS setup achieved comparable coverage in just 6~h, a 40$\%$ reduction in acquisition time. This improvement in efficiency is particularly valuable in operational railway environments, where minimizing track downtime is critical to avoid service disruptions.

\subsection*{Bayesian inference for decision support}
Bayesian approach in the inversion process provides a powerful framework for decision support in railway infrastructure diagnostics. Unlike deterministic approaches that yield a single ‘best-fit’ model, Bayesian inference characterizes the pdf of parameters, explicitly quantifying uncertainty. This probabilistic representation is particularly valuable in complex environments such as RT, where heterogeneities, degraded layers, and limited resolution can hinder confident interpretation.

By analysing the \textit{posterior} distributions of model parameters, such as $V_{s}$ or $H$, decision-makers gain access to critical information on the range and likelihood of plausible RT mechanical states. An important advantage of the Bayesian inversion approach lies in its ability to provide honest and transparent results by explicitly revealing the non-uniqueness of the solution. Indeed, the Bayesian framework captures the full range of plausible subsurface models through the \textit{posterior} \textit{pdf}, enabling the identification of regions where multiple solutions fit the data. Features such as bimodal or broad distributions indicate areas with limited resolution. The parameter $V_{s3}$, in particular, reveals spatial variations in resolution, with some profiles showing ambiguous results in the form of spread-out or bimodal \textit{pdfs}. These patterns reflect uncertainties in the inversion, suggesting that, in certain locations, the data do not sufficiently constrain $V_{s}$, possibly due to complex subsurface structures or limited resolution of the surface-wave data.

\section*{Conclusions}
\label{sec:conclusions}
%%%
Geophysical investigations were carried out along a section of the South-East European HSL where recurrent track geometry degradations due to mud pumping and RT anomalies, had been reported. Mechanical characterization of the RT and shallow RE using MASW revealed significant contrasts in $V_s$ profiles between two locations along the cess. To enable broader and more efficient deployment, a LS system was tested and validated against conventional seismic acquisition setup. Once validated, the LS was deployed directly on the track, demonstrating its suitability for the railway environment while reducing acquisition time. Despite increased noise due to interactions with the ballast and track superstructure, surface-wave analysis remained effective in detecting a LVL along the study site. This LVL, found to be thicker in areas with track degradation and confirmed by core samples. Together, these findings point to a localized structural disturbance within the shallow RT, directly aligned with observed surface degradation. Moving forward, probabilistic inversion techniques based on Bayesian inference, offer a pathway to transform surface-wave analyses from qualitative assessments into quantitative diagnostic tools. By explicitly incorporating uncertainty into the interpretation, this approach make it possible to develop maintenance strategies that are more reliable with regard to the mechanical state of RT.

While this study focused primarily on MASW, future work could greatly benefit from integrating GPR data during inversion process. In particular, GPR could provide valuable complementary information about dielectric property variations and stratigraphy, helping to reduce the non-unicity of the solution. A joint inversion approach combining GPR and MASW data would improve the resolution and reliability of $V_s$ models, especially by using GPR-derived constraints on the number and thickness of layers. Such an integrated geophysical strategy could enhance the diagnosis of RE conditions and improve the identification of structurally weakened zones.

\bibliographystyle{elsarticle-harv} 
\bibliography{biblio_tot}

\end{document}